# DATA SECURITY IN MOBILE DEVICES BY GEO LOCKING

M Prabu Kumar[1] and K Praneesh Kumar Yadav[2]

[1]Asst.Professor, School of Computing Sciences, VIT University, India
mprabukumar@vit.ac.in
[2]Graduate Student, School of Computing Sciences, VIT University, India
praneeshkumar@ymail.com

### ABSTRACT

*In this paper we present a way of hiding the data in mobile devices from being compromised. We use two level data hiding technique, where in its first level data is encrypted and stored in special records and the second level being a typical password protection scheme. The second level is for secure access of information from the device. In the first level, encryption of the data is done using the location coordinates as key. Location Coordinates are rounded up figures of longitude and latitude information. In the second phase the password entry differs from conventional schemes. Here we have used the patterns of traditional Rangoli for specifying the password and gaining access, thus minimising the chances of data leak in hostile situations. The proposed structure would be a better trade off in comparison with the previous models which use Bio Metric authentication – a relatively costly way of authentication.*

## 1. INTRODUCTION

The small size, relatively low cost and constant mobility of mobile phones make them invaluable for advocacy work but also make them more likely to be stolen, temporarily misplaced, lost or confiscated. Mobile phones carry a vast amount of data, which vary from contacts to logs of calls made and received and even SMS messages sent and received [8]. Security threats existing in mobile phone network could range from passively eavesdropping to actively stealing data from others. By virtue of carrying a list of all contacts mobile phone shows exactly with whom the user is working with. For users working on sensitive area this information may make both the user and everyone else in the network vulnerable.

Today, mobiles are replacing personal computers for most of the executives because mobiles are handy and the data is easily accessible without much of time for waiting to retrieve the data [8]. Accessing the information present on a device, especially from one that is in network is very easy. Hence, mobile phones are the next target for data criminals and as they possess the most sensitive information they are more vulnerable for attacks [9]. Encrypting the information in any form is still a viable for attackers. The mobile device is expected to have a special region where the under defined file storage structure is implemented. [3] In the current proposal information is encrypted and stored in different places with data at each record ending with a reference pointer to the record containing the continuation of the message.

## 2. Organisation of the Paper

The rest of the paper is organised as follows. Section 3 includes a short description and limitations of existing authentication models. Section 4 includes details about related work being carried out on securing data in mobile devices. Section 5 includes description on the records structure (proposed) for implementing the current model. Section 6 gives a bird's eye view of the proposed security model. Section 7 and Section 8 of the paper describe the encryption and decryption procedures respectively. Section 9 describes the password scheme to be implemented for imposing a higher degree of security over the content present in the device. Section 10 presents a short list of advantages of the current system in implementation. Section 11 concludes the paper with a brief description on the future work.

## 3. Existing Models

Many authentication schemes are available today, for validating a user to gain access to the concealed information. They can be broadly classified into Password Based Authentication Models and Bio Metric Authentication Models. Both these models have their own draw backs in terms of vulnerability and cost factors respectively.

Password based authentication models lack in making user comfortable with the system. As usually, lengthier the password – more secured the system is. These textual passwords with more number of special characters, numbers and alphabets in varying cases are difficult to remember and are more prone to attacks, if stored in insecure places [4]. Moreover accessing passwords at public places bear the risk of being stolen by any attacker observing the key strokes or shoulder surfing [4].

Biometric models are preferred to conventional password schemes as they overcome the above mentioned limitations of conventional password schemes. In spite, of their advantages over traditional schemes, bio metric models require relatively higher requirements in terms of resources, time and computation abilities [3]. Changing passwords when felt compromised is easier unlike in case of biometrics where in any one can copy user voice or digital biometric files/ databases and thus losing the accessible information forever [5].

These systems only provide security in terms of authentication i.e., direct access to the device only; but cannot protect the information when the device is hacked or stolen. The current model which stores the data all the time in encrypted format only; provides a better alternative to the above models.

## 4. Related Work

In the literature many models for securing data on mobile devices and PDA s are present. In [3], the security model involves both textual and bio metric authentication techniques for securing data in mobile devices. This model uses both local and remote verification modules for both textual and biometric systems. This model best fits for people using the devices in building/city but cannot cater to the needs of mobile users in global arena.

In [2], the data is hidden in binary text documents by embedding data in 8- connected boundary of a character. Conversions in these methods not only yield irregularities that are visually noticeable but also consume a lot of resources. As mobile devices work with limited resources this method of data hiding is infeasible.

In [6], a new approach to protect user's important data of mobile device from malicious activity is proposed which at the same time prevents the leakage of important by continuously running a monitoring service. This reduces power consumption by not inspecting all the

outgoing/ outbound traffic. This method if implemented releases the balloon of confidential contacts information, memos and calendar. Moreover, the user is forced to classify the information during storing, which is quite a cumbersome process. The current model overcomes the few limitations that the ongoing works carry.

## 5. System Overview

The section describes the overview of the proposed system. The following diagrams represent the view of messages that is shown to a user without proper privileges. All the texts including the details to be displayed are made to some junk and are presented.

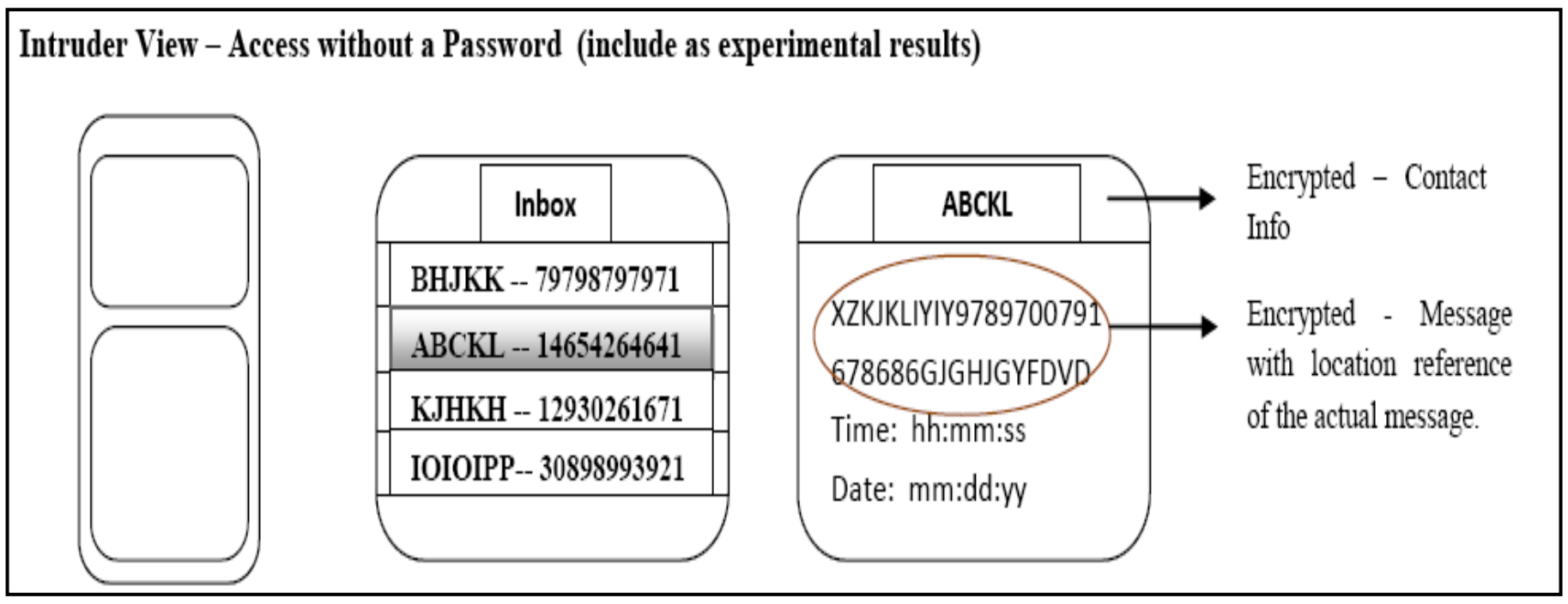

Fig 1. Intruder View – Access without a Password

Once a user with proper privileges gains access into the system, all the information required by the user is shown in user understandable format. The junk values that are displayed are generated using any one of the standard encryption algorithms.

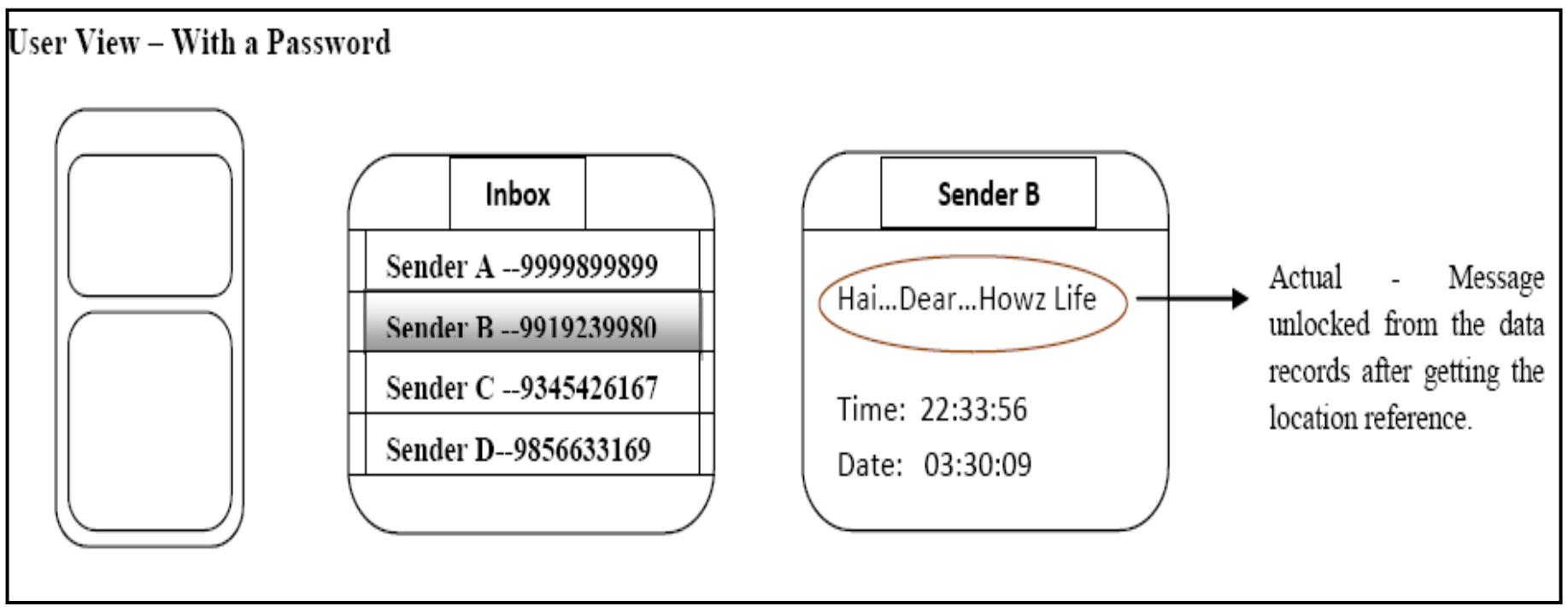

Fig 2. User View – Access with a Password

The following sections describe the nature of storage and storage structures proposed for this type of security.

## 6. Data Storage

The data to be stored, considering a text message or a memo or a reminder is assumed to be in record form – as like as a key, value pair, where the location is the key and the values in that location are the information stored in it.

Table 1. Record View – Empty Key Value Pairs

| Hash (or) Address | Encrypted Value of Text (can be a memo/message/ |
|---|---|
| | |
| | |
| | |
| | |
| | |
| | |
| | |

Every key value pair is stored as in *fig 2*. The content to be stored is usually split into many pieces depending on the size. A typical text to be stored as a value can be of length 150 characters. Any text exceeding the size is broken into parts of length 150 chars, encrypted using transposition ciphering technique and stored. The key description is included in the ***section 3.1***.

The encrypted value of the text contains a delimiter, indicating the next memory location or the end of the message. Any value at the end with a number other than '**0**' indicates the address of the continued location. The number 0 indicates the end of the message.

Table 2 .Record View – Filled Key Value Pairs

| Hash (or) Address | Encrypted Value of Text (can be a memo/message/ |
|---|---|
| 1 | Encrypted (We are working on well....things are going on 8) |
| 2 | Encrypted (Hai...Dear...Howz Life 0) |
| 3 | Encrypted (Dinner at 3:00 pm City Center.... try to make up 0) |
| ..... | ........................................................................................ |
| ...... | ........................................................................................ |
| 8 | Encrypted (very well...shall meet Joe on 20 March 1800 hrs 0) |
| ....... | ........................................................................................ |

The data that is being stored in the mobile can be scrambled and stored in two various records. The scrambled text can be stored after encrypting it. The explanation is by taking only the contact information into consideration. This technique can be extended to all types of information that are being stored in the mobile phone. The following block diagram illustrates the encryption procedure that can be followed for encrypting and storing.

## 7. ENCRYPTION

The encryption of the data that has to be stored is done in various steps starting from data scrambling and ending with saving into the records (***section 2***). The following discussion uses a text memo that has a name of a person (Contact Name) and the contact number pertaining to the person (Contact Info. say a 9 digit number).

**Step 1 - Combining**: In this step the whole of the data (text) that is received is made a single line of text. Here the contact name i.e., PQRST and the corresponding contact information i.e., 976543767 are both padded with each other and the memo is made PQRST 976543767, separated by a delimiter here a blank space.

**Step 2 – Jumbling & Splitting**: Here the text that is added is scrambled in a predefined sequence and the jumbled sequence is then split into two parts, preferably equal sizes. Hence the text after scrambling becomes P6SQ9767R73S4T and split into two parts (say A1 & A2), where A1 - P65Q9767 and A2 – R73S4T.

**The Cipher**:

```
for(n=0;n<2;n++)
{
k=0;
for(i=0;i<new11/blk;i++)
{
for(l=0;l<blk;l++)
{
cipher[k++]=string[key[l]-1+i*blk]; //Encryption
}
}
for(m=0;cipher[m]!='\0';m++)
string[m]=cipher[m];
}
cipher[k]='\0';
```

**Step 3 – Key Generation**: Key generation is the most important part of the process in storing the text content in a mobile device. Assume the device is at 26.15875768° Latitude and 32.153457537° Longitude position at the time of receiving the message or storing the information on the device. Only two digits of decimal precision are considered to reduce massive floating point operations in the device. Hence the latitude value that is taken will be 26.15° and the longitude value taken to be 32.15°.

The key pair is generated as follows:

**Key 1:** 2615 + Padding which includes numbers other than the numbers present in the Key1 (i.e., other than 2, 6, 1 & 5) and that are less than the greatest number in the key1 are 3 & 4. Hence the Key 1 becomes 261534.

**Key 2:** 3215 + Padding ( as mentioned above) but here the maximum number is 5 and there exists only one number that is other than 1, 2, 3 & 5 which is even <5. Both the keys can't be of unequal size. Hence the maximum number available (here 5) is incremented by 1 i.e., it is made 6 and is padded along with previously generated number i.e., 4. Hence the key 2 become 321564.

**Step 4 – Encryption *(applying the Transposition Cipher)*:** The encryption of the messages after the key generation is done using the Transposition Cipher Technique. After the encryption the key A1 becomes 7Q65P9zz7z6z, say EA1 and the key A2 becomes R73S4T, say EA2.

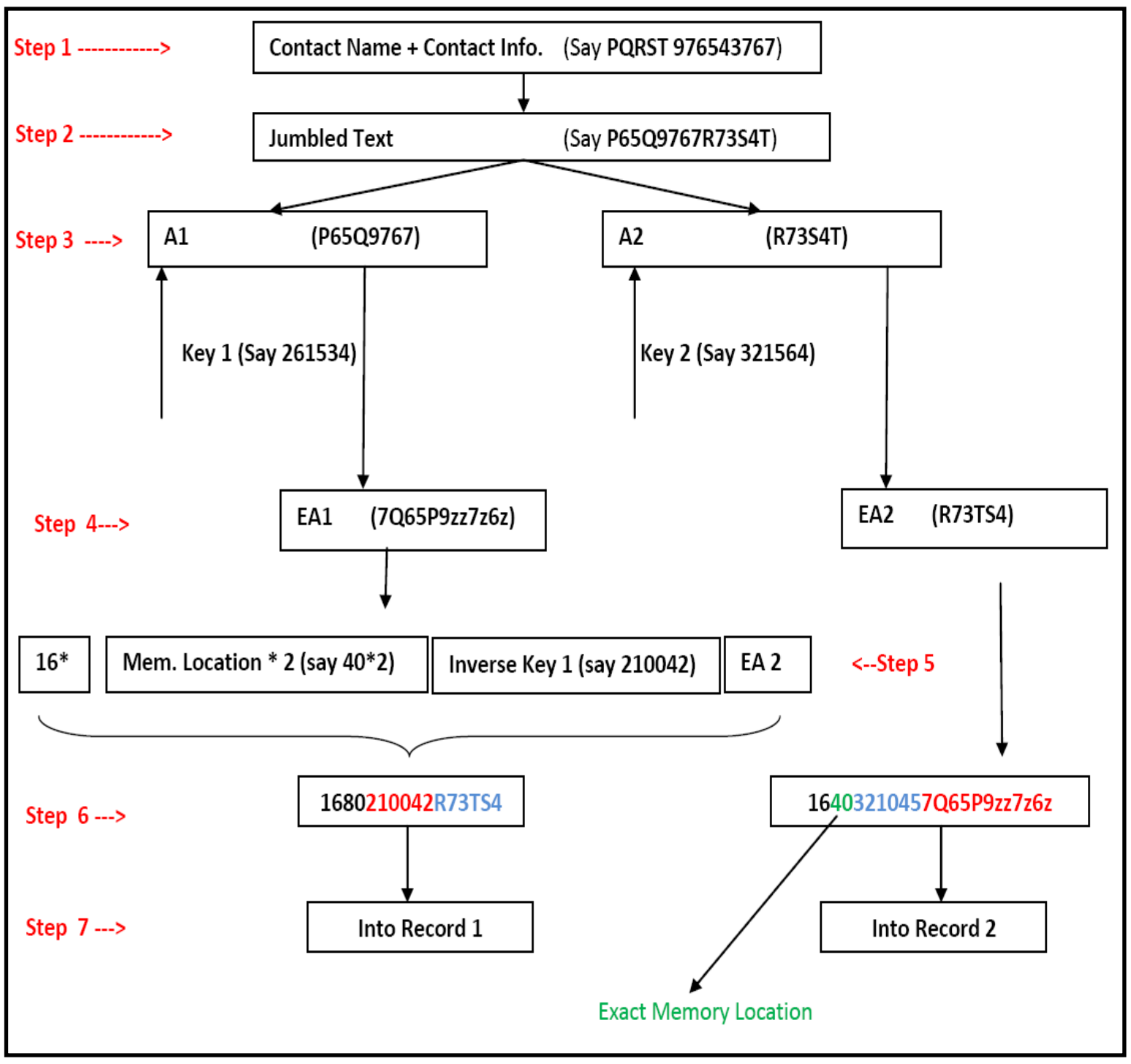

Fig 3. Encryption – Block Diagram

**Step 5 & Step 6– Padding:** In this stage the encrypted data are jumbled and are padded with different sequences i.e., the first part of the encrypted message is padded with inverse key of the key that is used for encrypting the second part of the message (A2). A number representing the starting letter of the message (here 16, since the text content started with P) is even padded to the combined pad of Inverse Key of K1 and EA2. Twice the value of hash address even is padded to the encrypted message.

Thus the first part of the saved text is – 1680210042R73TS4, the second part of the same is – 16403210457Q65P9zz7z6z -

where ABCD

A – 16, since P is the 16$^{th}$ letter in alphabets
B – 2 times the location in first part and the actual location in second part
C - Inverse Keys of K1 and K2 respectively
D – Encrypted text halves

The following fig -5 implements the generation of the Inverse Keys with a preset block size

```
if(dbpwd.length()!=0)
{
int key11[]=new int[20];
int b1=23415;//Input Key
int rem11=0;
int length11=5;//Lenghth of Input Key
int inv[]=new int[20];
int blk2=5;        //Block Size
for(i=0;i<length11;i++)
{
key11[length11-i-1]=b1%10;
rem11=b1/10;
b1=rem11;
}
int l1=0,mno=0;
for(int i1=0;i1<blk2;i1++)
{
if(i1+1==key11[l1])
{
inv[mno]=i1+1;
("The Inverting Key111"+inv[mno]);
mno=mno+1;
l1=l1+1;
}
else
{
l1=l1;
}
}
```

The Inverse Key Generation – Java Implementation Code

**Step 7 – Storage:** Now both these parts of the text content say H1 & H2 are stored as key value pairs in the records (section 2), where H1 - 1680210042R73TS4 and H2-16403210457Q65P9zz7z6z. During storing care is taken that the header information (information other than the $EA_i$) are separated from the encrypted data. An extra delimiter is added if the 2nd part of the $H_i$'s (i.e., the address) is of 2 digits. The preferred header length is 11 characters thus enabling to store 498 (i.e, 999 unique identities for address and half of it is 498 for each equal sized record set) messages or memos etc. The $H_i$ s during storing will be H1-1640321045*7Q65P9zz7z6z and H2 – 1680210042*R73TS4.

***Storage Header***

| Character Information (2 Chars) | Position Information (up to 3 Chars) | Inverse Keys (6 Chars) | Encrypted Text |
|---|---|---|---|

Fig 4. The Storage Header

## 8. Decryption

**Step 1 Retrieval**: All the pieces of information for a particular data that are stored in various records are collected before decrypting them to form the actual content.

**Step 2 Separation**: The scrambled and encrypted data from each record is now separated for the actual record location, the inverse key and the encrypted text information.

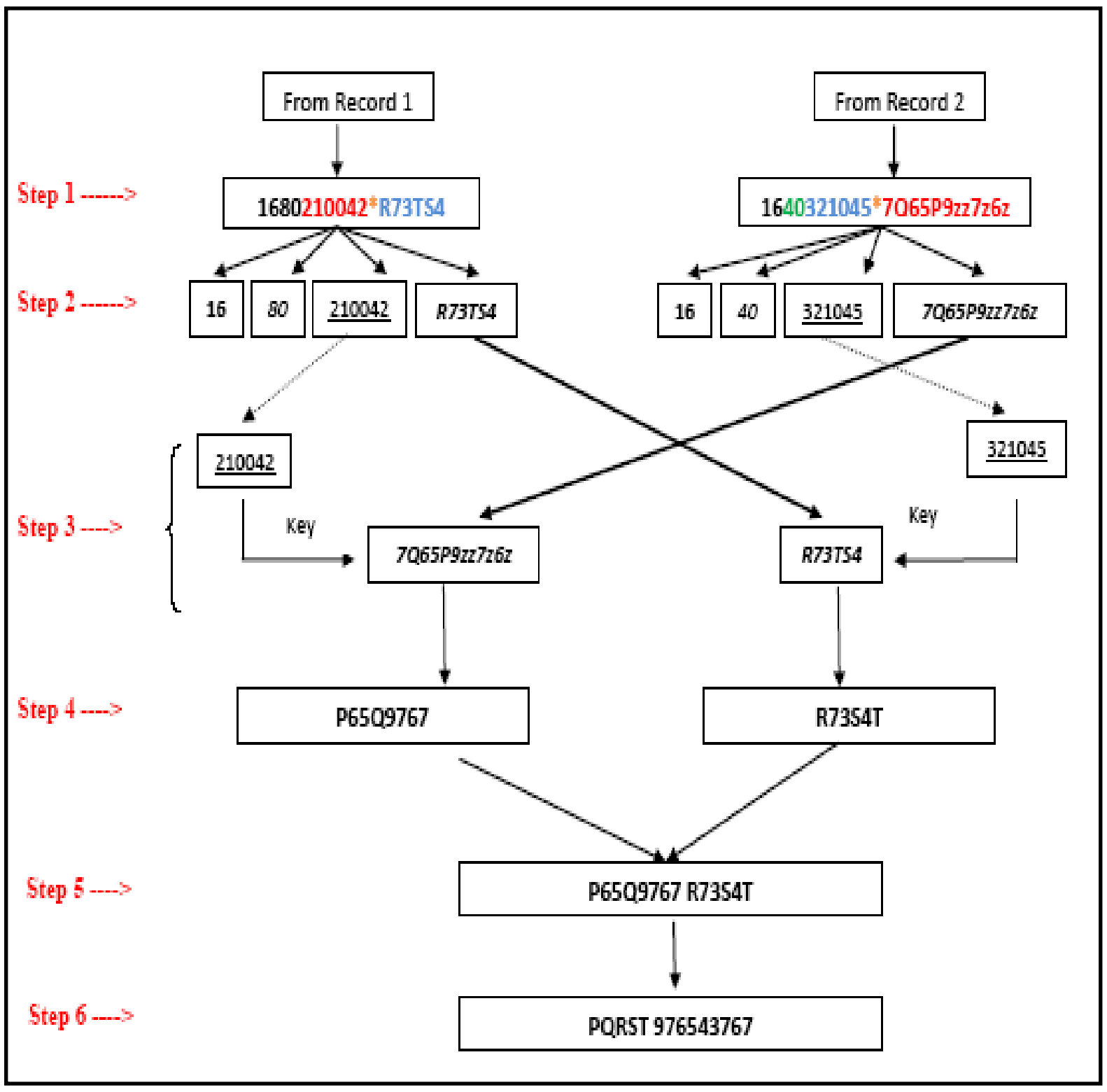

Fig 5. Decryption – Block Diagram

**Step 3 Exchange**: The encrypted texts from both the records are exchanged. This step brings the actual encrypted text to its corresponding inverse key.

**Step 4 Decryption**: The decryption of the available text is carried on using the same algorithm (used for encryption, here Transposition Cipher) and the parts of the scrambled texts are made available.

**Step 5 Concatenation**: The pieces of scrambled texts are now concatenated so as to make the actual scrambled text which is passed to next phase for generating the actual text message.

**Step 6 Inverse Permutation**: The scrambled text is now re permuted to derive the actual text message. This procedure is followed for every piece of information that is stored on the device.

## 9. Passwords

The data locking feature of the mobile phone is an inspired output from the traditional Indian *Rangolis*. Rangolis are formed by joining the few pre marked points in a skilled manner. The current algorithm discusses the way access can be granted to a user into his mobile data.

The user has to select the level of security he requires. The following figure explains the manner in which one can gain access to his account.

In the Fig 8, the password pattern is selected by the user in a matrix of size 4 * 4. Innumerable patterns can be generated from the set of points. Greater the size of the matrix, lesser is the probability of password being cracked.

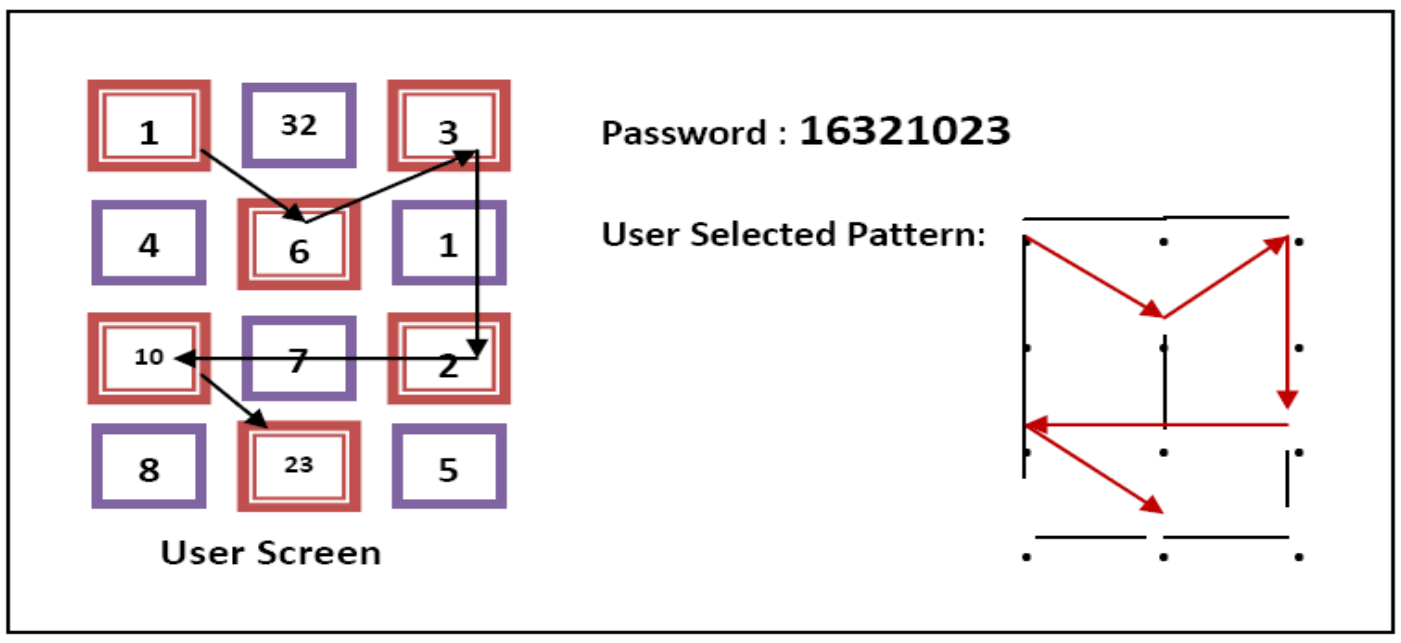

Fig 6 The Password Input Screen

In the Fig 8, the password pattern is selected by the user in a matrix of size 4 * 4. Innumerable patterns can be generated from the set of points. Greater the size of the matrix, lesser is the probability of password being cracked.

The user has to type the password in the preselected pattern only. The password changes once the user crosses his predefined boundary, hence reducing the chances of data being stolen in case of mobile confiscation or loosing. New passwords are generated on the lines of GPS location only. The user screen for password prompting changes every time the new password is generated. The new password can be activated only with the help of existing password and user can even skip this step when prompted for.

The message store/ memo/ calendar can be even accessed without passing the above step, but the intruder can see only the decrypted text of Record1 (figure 3) only. Even after entering into the phone book, the intended user cannot see all the contacts at the same time. Only the desired set of information can be seen. This protects all the information from being exposed at the same time.

## 11. Advantages

- As the pattern is only known to user, accessing the information on the mobile phone becomes quite difficult even the password is known.
- As the information is being locked on the basis of location, it is difficult to find out the similarity in the encrypted texts.
- As the password changes once the user crosses the fixed area, it is highly impossible to gain access into the information.

## 12. Conclusion

In this paper, a new data hiding technique for information on mobile devices is presented. This technique imposes a two level data hiding mechanism, where in the first level implements the storing techniques and the second level implements the accessing mechanisms. The use of a predefined pattern for password entry will enhance the security of data in case of device theft or during hostile situations. The use of multiple records for saving the parts of a message protects the data from being stolen in case of device being hacked. The future enhancements would be concentrating on imposing a secured check sum for the outbound data, for the third level of data hiding.

**Authors**

**Prabukumar.M** was born in Salem city, Tamilnadu, India on July 25, 1981. He obtained his Bachelor of Engineering in Electronics and Communication Engineering from Periyar University, Salem during 1998 - 2002. He did his Master of Technology in Computer Vision and Image Processing from Amrita Deemed University, Coimbatore. He is currently working as Asst.Professor in School of Computing Sciences of VIT University, Vellore, India. He is life time member in Computer Society of India. His areas of interests include image processing, algorithms, VLSI Design of image processing algorithms, Video Image Processing, Computer Vision, Digital Signal Processing and Digital Systems Design, Computer Networks.

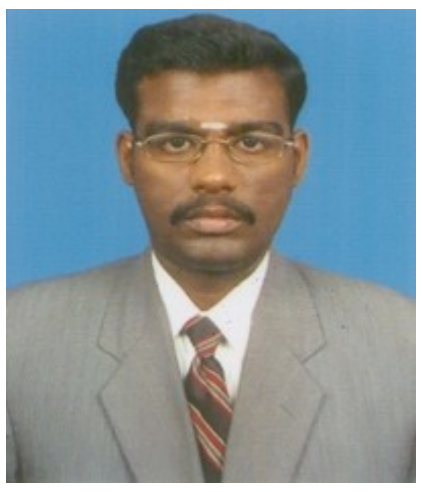

**K Praneesh Kumar Yadav** was born in Andhra Pradesh, India on June 22, 1988. He is currently pursuing his Masters in Software Engg. from Software Engg. Division of School of Computing Sciences, VIT University. He is a student member of Computer Society of India and currently the head of the Students Branch of the same at VIT University. He served as the Campus Ambassador for Sun Microsystems. His areas of intrest include development of security and automation systems.

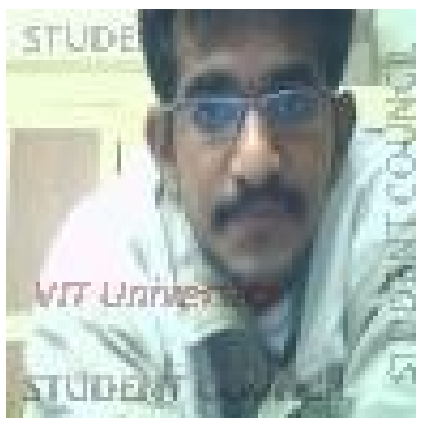